\documentclass[twocolumn,floatfix]{revtex4} 
\usepackage{ifthen} 
\usepackage{color}
\usepackage{umlaut}
\usepackage{amsmath}
\usepackage{amsfonts}
\usepackage{graphicx}%
\usepackage{dcolumn} 
\newboolean{offen}
\setboolean{offen}{false}
\newcommand{\geheim}[1]{\ifthenelse{\boolean{offen}}{#1}{}}

 \newcommand{\ket}[1]{{| #1\rangle}}

\newcommand{\braket}[2]{\langle #1 | #2 \rangle}
\newcommand{\matel}[3]{\langle #1 | #2 | #3 \rangle}

\begin{document}

\title{Spin Chains as Perfect Quantum State Mirrors}

\author{Peter Karbach and Joachim Stolze} 
\affiliation{Institut für Physik, Universität Dortmund, 44221 Dortmund, Germany}

\date{\today}

\begin{abstract}
  Quantum information transfer is an important part of quantum information
  processing. Several proposals for quantum information transfer along
  linear arrays of nearest-neighbor coupled qubits or spins were
  made recently.  
Perfect transfer was shown to exist in two models with specifically designed
strongly inhomogeneous couplings.
We
  show that perfect transfer occurs in an entire class of
  chains, including systems whose nearest-neighbor couplings vary only
  weakly along the chain. The key to these observations
  is the Jordan-Wigner mapping of spins to
  noninteracting lattice fermions which display perfectly periodic
  dynamics if the single-particle energy spectrum
is appropriate. 
After a half-period of that dynamics any state is
transformed into its mirror image with respect to the center of the chain.
The absence of fermion interactions
  preserves these features at arbitrary temperature
  and allows for the transfer of nontrivially entangled states of
  several spins or qubits.
\end{abstract}


\maketitle


Quantum information processing (QIP) \cite{NC01} has been an
increasingly important area of physics research over the past decade.
The generic building block of QIP is the qubit, which is naturally
realized as a spin-1/2 particle. A multitude of coupled spin-1/2
systems have been discussed as possible candidates for the quantum
gates needed in quantum computing 
\cite{Bur04}. Only recently a new
focal field of activity has developed, dealing not with the
processing, but with the {\em transport} of quantum information. As most
kinds of directed transport 
take advantage of one-dimensional structures it seems natural
to explore the possibilities of one-dimensional arrays of coupled
spin-1/2 systems as transmission lines for quantum information. 

In \cite{KG02} a sequence of external RF pulses was proposed to drive
single-spin quantum information down
a chain of
Ising-coupled spins.  Other studies
proposed to use the natural internal dynamics of coupled spins for the
transfer of information.  In a homogeneous ferromagnetic Heisenberg
chain \cite{Bos02} initially in its ground state, a
single-spin state generated at one end of the chain is transferred
to the other end with reasonable (but not perfect) fidelity by means
of spin waves. This approach is restricted to zero temperature and
single spin-wave (and consequently single spin-flip) states since
multiple spin-wave states are unstable under the Heisenberg
interaction. This excludes the transport of
entanglement, except for states of the type
$\alpha \ket{\uparrow \downarrow} + \beta \ket{\downarrow \uparrow}$
(with the two spins initially located at fixed sites). The time evolution
of these states was studied analytically \cite{Sub04} in an otherwise
completely polarized infinitely long ferromagnetic Heisenberg chain.
Initial states $\alpha \ket{\uparrow \uparrow} + \beta \ket{\downarrow
  \downarrow}$ were also studied analytically but with the Heisenberg
interaction changed to an XX interaction in order to exclude
spin-wave interactions.  Spin chain models for
single-spin quantum information transport may be implemented as
Josephson junction arrays \cite{RFB05}; another proposal \cite{PPKF05} 
involves more general spin networks.

Gaussian spin wave packets (i.e. Gaussian
weighted superpositions of single-spin-flip states) may be more
suitable for information transfer than single localized flipped spin
states \cite{OL04}, because they  may be tailored so
as to occupy only the least dispersive part of the spin-wave
dispersion relation $\omega(k)$.  Two \cite{BB04,BB05} or more
\cite{BGB04} spin chains may be used in parallel to enhance the
fidelity of quantum information transfer via appropriate protocols.
The dynamics of several entanglement-related quantities in the ground
state of an infinitely long anisotropic XY chain with a supercritical
magnetic field were also studied exhaustively \cite{AOPx03},
confirming the typical power-law decays found earlier in
correlations of this model and related ones
\cite{Nie67,McC68,BM71,MBA71,VT78,MPS83a,MPS83b,MS84}.

Other studies have looked for cases where a
quantum state is transported {\em
  perfectly}, that is, without any decay or dissipation, along a spin
chain. Perfect transmission is possible in a translationally invariant
ring, if the time evolution operator for some particular time is equal
to a lattice translation; it was found, however, that the
twelve-site antiferromagnetic Heisenberg ring does not fulfill the
necessary conditions \cite{SS04}. Perfect transmission 
was demonstrated \cite{CDEL04}
for an open-ended XX chain with {\em inhomogeneous} coupling:
the amplitude for the transfer of a single flipped spin from one end
of a completely spin-polarized chain to the opposite end is unity
for certain times. This result is understood intuitively by
observing that the, say, $2J+1$ sites of the chain can be mapped to
the $2J+1$ eigenstates of $J_z$ for a single fictitious particle with angular
momentum quantum number $J$. With appropriately chosen matrix elements
 along the spin chain the motion of the single flipped spin from one
 end of the chain to the opposite end, and back, corresponds to the
 motion of the fictitious particle's angular momentum in a transverse
 ($x$ or $y$) magnetic field. That motion brings the fictitious
 particle from the $J_z=J$ state to the $J_z=-J$ state and back
 periodically. 
Perfect ground-state transport of more general states is also possible
\cite{ACDE04}. A state involving a number of spins
in the left half of a symmetric quantum
spin chain be {\em mirrored} about the middle of the chain so that the local information contained in it is effectively
transported to the right half of the chain. 
The inhomogeneous XX chain of 
\cite{CDEL04} shares this property with a similar system possessing an additional
site-dependent magnetic field in $z$ direction.
 
Here we analyze the ingredients necessary for perfect quantum state
mirroring (for both pure and mixed states) and we show how the two
particular systems described in \cite{ACDE04} can be generalized to an
infinity of cases. This opens up possibilities for quantum spin chain
engineering which may be used for the design of spin chains with
desirable properties in addition to perfect state transfer. As an
example we construct a spin chain without local magnetic fields and
with only mildly inhomogeneous couplings. In fact the necessary degree
of inhomogeneity decreases with growing chain length.  We further show
that the perfect mirroring property is stable against thermal
fluctuations and we demonstrate how that property makes arbitrary spin
autocorrelation functions perfectly periodic in time.

The model we consider is an inhomogeneous open-ended
$(N+1)$-site  $S=\frac{1}{2}$ XX chain with Hamiltonian
\begin{equation}
  \label{1}
  H=2 \sum_{i=1}^{N} J_i (S_i^x S_{i-1}^x +S_i^y S_{i-1}^y)  +
\sum_{i=0}^{N} h_i \left( S_i^z + \frac{1}{2} \right),
\end{equation}
where $S_i^{\alpha} (\alpha=x,y,z)$ are the usual $S=\frac{1}{2}$
operators with eigenvalues $\pm\frac{1}{2}$, $J_i$ and $h_i$ are
local couplings and magnetic fields, respectively.
Eq. (\ref{1}) can be mapped to a Hamiltonian of noninteracting 
spinless lattice fermions,
\begin{equation}
  \label{2}
  {H}=  \sum_{i=1}^{N} J_i (c_{i-1}^{\dag} c_{i} +
  c_{i}^{\dag} c_{i-1}) +  \sum_{i=0}^{N} h_i c_{i}^{\dag} c_{i}
\end{equation}
by means of the Jordan-Wigner transformation \cite{LSM61,Kat62} between spin
and fermion operators:
\begin{equation}
  \label{3}
  S_i^z = c_i^{\dag} c_i - \frac{1}{2},
\end{equation}
\begin{equation}
  \label{4}
  S_i^+ = (-1)^{\sum_{k=0}^{i-1}c_k^{\dag} c_k} \; c_i^{\dag} = \prod_{k=0}^{i-1}
  (1 - 2 c_k^{\dag} c_k) c_i^{\dag}.
\end{equation}
Due to its bilinear nature Eq. (\ref{2}) can be diagonalized,
\begin{equation}
  \label{5}
  H = \sum_{\nu=0}^N \varepsilon_{\nu} c_{\nu}^{\dag} c_{\nu}.
\end{equation}
Here $c_{\nu}^{\dag}$ creates a fermion in a single-particle
eigenstate of energy 
$\varepsilon_{\nu}$, whereas $c_i^{\dag}$ creates one at lattice site
$i$.  Once the $\varepsilon_{\nu}$ and the corresponding
eigenstates are known, the dynamics generated by $H$  can be
calculated in detail, since every eigenstate of $H$ is uniquely
characterized by the fermion occupation numbers of the single-particle
states.

The single-particle energies $\varepsilon_{\nu}$ $(\nu=0,...,N)$ and
the corresponding eigenstates are the eigenvalues and eigenvectors,
respectively, of a symmetric tridiagonal matrix $H_1$ with diagonal
elements $(h_0,h_1,...,h_N)$ and subdiagonal elements
$(J_1,J_2,...,J_N)$; all $J_i$ are strictly positive \cite{negative}.

We further assume the system to possess a mirror symmetry, 
$h_i=h_{N-i}$ and $J_i=J_{N+1-i}$. The
$(N+1)$-dimensional eigenvectors $\vec{x}$ of $H_1$, 
$H_1 \vec{x}= \varepsilon_{\nu} \vec{x}$, then have definite parity, that is,
every eigenvector is either even, with components $x_i=x_{N-i}$, or odd,
$x_i=-x_{N-i}$. This resembles closely the situation in elementary
one-dimensional (continuum) quantum mechanics, where for a symmetric
potential the energy eigenfunctions have definite parity and are
nondegenerate. From the well-known theorem relating the number of
zeros of the wavefunction to the number of the energy eigenvalue (in
ascending order) one can then conclude that even and odd
eigenfunctions alternate as energy increases. It turns out that the
same line of argument is possible for the quantum lattice system of
interest here,
due to the following theorem \cite{GK02}: 
{\em 
For a real symmetric tridiagonal
matrix with only positive subdiagonal elements (i) all eigenvalues are
real and nondegenerate, and (ii) the sequence of the components of the
$j$th eigenvector (in ascending order of the eigenvalues, $j=0,1,...$)
shows exactly $j$ sign changes.}
This implies that the eigenvectors of $H_1$ are alternately even and
odd. 
To achieve perfect mirror inversion of an arbitrary {\em
  many-particle} state by some operation $M$ it is sufficient that $M$
transforms all {\em single-particle} eigenstates into their mirror
images. Note that $M$ is not simply the reflection (parity) $P$, since
$P$ changes the sign of all odd single-particle states. That sign
change must be compensated for 
by a dynamic phase factor $\exp [i \pi (2n+1)]$ ($n$ an arbitrary
integer)
 to achieve perfect
mirror inversion. 
This entails conditions on
 the energy spectrum $\varepsilon_{\nu}$:
a single-particle state $\ket{i}$ localized at site $i$ evolves in time
according to 
\begin{equation}
  \label{eq:1}
  e^{-iHt} \ket{i} = \sum_{\nu=0}^N e^{-i \varepsilon_{\nu}t}
    \ket{\nu}\braket{\nu}{i}, 
\end{equation}
where $\ket{\nu}$ is a single-particle energy eigenstate with energy
$\varepsilon_{\nu}$ ($\hbar=1$). The alternating
parity implies $\braket{N-i}{\nu}=(-1)^{\nu}
\braket{i}{\nu}$ and thus 
\begin{equation}
  \label{eq:2}
  \ket{N-i} = \sum_{\nu} \ket{\nu}\braket{\nu}{N-i}= \sum_{\nu}
  (-1)^{\nu} \ket{\nu}\braket{\nu}{i}.
\end{equation}
Perfect quantum state mirroring occurs if, for some time $\tau$ time evolution equals reflection,
\begin{equation}
  \label{eq:3}
  e^{-iH\tau} \ket{i} = e^{i \phi_0} \ket{N-i},
\end{equation}
(up to some global phase $\phi_0$) for all $i$. This is obviously the case if
$e^{-i\varepsilon_{\nu} \tau} = e^{-i (\pi \nu + \phi_0)}$, or
equivalently
\begin{equation}
  \label{eq:4}
  \varepsilon_{\nu} \tau = (2 n(\nu) + \nu)\pi + \phi_0,
\end{equation}
where $n(\nu)$ is an {\em arbitrary} integer function.
Note that 
every spatially symmetric  system whose
single-particle energies obey (\ref{eq:4}) generates a perfect mirror
image of {\em any} input state since  it does so for all single-particle states.

The two systems discussed in \cite{ACDE04} correspond to (i) a linear
spectrum, $\varepsilon_{\nu}= \omega_0 + \nu \omega$, which, for
$\tau=\pi/\omega$, leads to $\phi_0=\pi \omega_0/\omega$ and $n(\nu)
\equiv 0$, and (ii) a quadratic spectrum, $\varepsilon_{\nu}= \omega_0
+ \nu (\nu + 1 + \frac{2p+1}{q}) \omega$, $p$ and $q$ positive integers,
which, for $\tau=q \pi/\omega$, leads to $\phi_0=\pi q \omega_0 /
\omega$ and $n(\nu)=q \frac{\nu(\nu+1)}{2} + p \nu$. 

As the function $n(\nu)$ in (\ref{eq:4}) is completely arbitrary there
are infinitely many single-particle spectra suitable for quantum state
mirroring.  For a given nondegenerate single-particle spectrum there
exists a unique symmetric tridiagonal Hamiltonian matrix with
nonnegative subdiagonal elements and with the additional spatial
symmetry properties discussed above \cite{Hal76}.  The actual
construction of that matrix from its eigenvalues may proceed either via a direct
algorithm \cite{Hoc74} or by
simulated annealing \cite{PTVF92}.
 The possibility of generalizing the work of
\cite{ACDE04} by solving an inverse eigenvalue problem was
already pointed out in \cite{YB04,SLSS04}.

While most schemes for quantum information transfer in spin chains are
restricted to states generated from the completely polarized ground
state by manipulating a single spin we stress that the class of models
discussed here perfectly mirrors states involving an arbitrary number
of lattice sites. This is due to the fact that the Jordan-Wigner
transformation (\ref{3},\ref{4}) maps the spins to {\em noninteracting}
fermions. Since due to (\ref{eq:4}) all single-fermion states are mirrored perfectly at
the same instant of time $\tau$, so are all many-fermion states, pure
or mixed, including finite-temperature density operators.

\begin{figure}[h]
\includegraphics[width=\columnwidth]{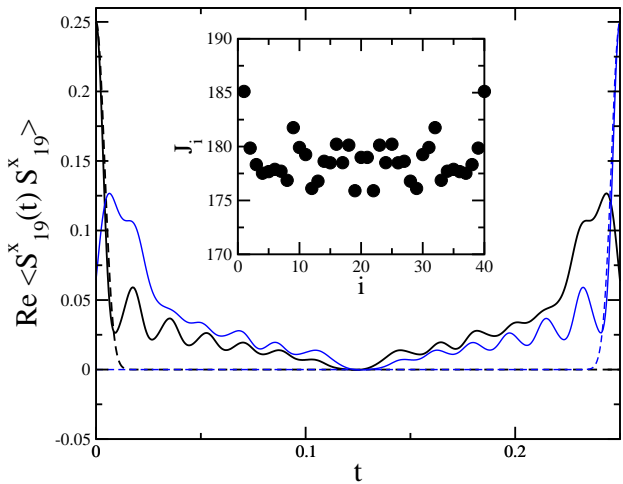}
\caption{ $x$ spin autocorrelation function (real part) at site 19
  of a 41-spin chain  at
  temperatures $T=0$ (solid line) and $T=10^4$ (dashed line). The
  maximum at the left edge of the graph is the one at $t=0$, that at
  the right edge (slightly thinner blue lines) is the one at
  $t=48\pi$, after 24 periods of the correlation function ($t$ axis shifted).
Inset: nearest neighbor exchange constants $J_i$.
}
\label{zwei}
\end{figure}

Since mirroring twice reproduces the initial state, the time evolution
of the system should be periodic with period $2 \tau$. This can in
fact be verified  by considering the equilibrium time autocorrelation
function of an arbitrary observable $A=A^{\dag}$:
\begin{eqnarray}
  \label{eq:5}
  \langle A(t)A \rangle &=& Z^{-1} \sum_n \matel{n}
{e^{-\beta H} e^{iHt} A e^{-iHt} A}
{n}\\ \nonumber
&=& Z^{-1} \sum_{n,m} e^{-\beta E_n} e^{i(E_n-E_m)t} 
|\matel{n}{A}{m}|^2 ,
\end{eqnarray}
where $Z=\sum_n e^{-\beta E_n}$ with $\beta=(k_B T)^{-1}$ is the canonical
partition function; $\ket{n}$ is a many-particle eigenstate of
$H$. Since all $(E_n-E_m)$ are  multiples of some energy,  $\langle
A(t)A \rangle$ is a periodic function of $t$. See Fig. \ref{zwei} for an example.

The freedom in the choice of $n(\nu)$ in (\ref{eq:4}) may be used for
{\em quantum spin chain engineering}, that is, the design of systems
with desirable additional properties. As an example of these possibilities we
show how the well-studied homogeneous XX chain ($J_i \equiv J$ in
(\ref{1})) may be modified to display the perfect mirroring property.
The system proposed in \cite{CDEL04} has $J_i$ growing from the
boundary to the center of the system, with a maximum value
proportional to the chain length, if the time for perfect transfer is
kept constant. If the maximum achievable $J_i$ is fixed by physical
restrictions, the $J_i$ towards the boundary of the system decrease $
\sim N^{-\frac{1}{2}}$ as the chain length $N$ grows and the
transmission time grows $\sim N$, implying a constant signal
transmission velocity \cite{CDDx04}. The very weak couplings close to
the ends of the chain would make such a system extremely susceptible
to external perturbations.  It is therefore interesting to know
whether the same performance (constant signal velocity) can be
achieved with more homogeneous ``wires'', that is, less variation in
the $J_i$. Note that $J_i \equiv J$ leads to the familiar spin wave
dispersion relation 
$\varepsilon_{\nu}= \omega(k)= 2J \cos k$ ($k=\pi(\nu+1)/(N+2)$)
for the single-particle states. For small $k$ thus $\varepsilon_{\nu}- 2J
\sim -k^2$ which fulfills the spectral condition (\ref{eq:4})
for perfect mirror inversion. It is possible, by only slightly
distorting the cosine dispersion, to fulfill that condition througout
the entire $k$ range. If the unit of energy (say, the energy gap above
the lowest single-particle eigenvalue) is  kept constant the
coupling  obviously scales with chain length as $J \sim N^2$. In
that case the transmission time is constant regardless of the chain
length.  For easy comparison of energy eigenvalues and periodicities
between chains of different lengths we decided to choose that
possibility. The eigenvalues $\varepsilon_{\nu}$ of
a chain with 31 sites, for example, may be chosen as
$\varepsilon_{\nu} = 
     0$
,$\pm    21$
,$\pm    40$
,$\pm    61$
,$\pm    80$
,$\pm    97$
,$\pm   116$
,$\pm   131$
,$\pm   146$
,$\pm   161$
,$\pm   172$
,$\pm   183$
,$\pm   192$
,$\pm   199$
,$\pm   204$
,$\pm   207$. The corresponding nearest-neighbor couplings $J_i$ then
vary between 101.5 and 108.5, a relative variation of $\pm 3.3$
percent. With growing length the variation in the $J_i$ decreases; 
for a 50-site system, for example, it amounts to 
only about one percent.
\begin{figure}[h]
\includegraphics[width=\columnwidth]{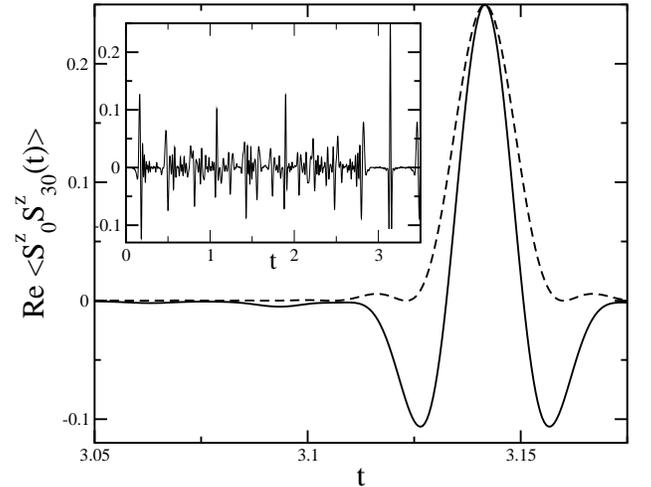}
\caption{ $z$ spin correlation function (real part) between the two
  ends of a 31-spin chain  at
  temperatures $T=0$ (solid line) and $T=1000$ (dashed line), in the
  vicinity of $t=\pi$. The
maximum possible value 1/4 of the
  correlation  at  $t=\pi$ demonstrates perfect state
  transfer. Inset: same correlation for $T=0$ over an extended time
  range. (Note the symmetry with respect to $t=\pi$; the correlation
  function is periodic with period $2\pi$.)
}
\label{eins}
\end{figure}

In Fig.\ref{eins} we demonstrate the perfect transfer of a single-site
state between the first and last sites of a chain with $N=30$.  The
correlation $\langle S_0^z(0) S_{30}^z(t) \rangle$ is zero at $t=0$
and reaches its maximum possible value 1/4 at $t=\tau= \pi$. Thus, an
$S^z$ eigenstate initially prepared at site 0 can be retrieved at the
initially uncorrelated site 30 after time $\tau$ at any temperature.
The inset of Fig.\ref{eins} gives an impression of the irregular
behavior of the correlation during the whole time range between 0 and
$\pi$.

Fig. \ref{zwei} shows the autocorrelation function of the $x$ spin
component at site 19 in a 41-site chain, demonstrating the
periodicity, eq. (\ref{eq:5}). It is important to
note that by the
Jordan-Wigner transformation (\ref{4}) a two-spin correlation
$\langle S_k^x(0) S_{k}^x(t) \rangle$
 becomes (in fermionic terms) a
complicated many-particle correlation involving lattice sites 0
through $k$. 
The single-fermion eigenstates and eigenvalues enter the correlation
function in an entirely nontrivial way; nevertheless it is perfectly
periodic. 
The figure shows the initial decay of the correlation at
$t=0$ and its 24th ``revival'' at $t=48 \pi$.  Note the
rapid decay and the absence of oscillations at high $T$. In fact, for
the homogeneous XX chain ($J_i \equiv J$) the $x$ autocorrelation is
known \cite{SJL75,BJ76,CP77} to be a Gaussian at $T=\infty$ in the
thermodynamic limit, while all nonlocal $x$  correlations vanish
identically. The inset shows the
weakly varying nearest-neighbor exchange constants $J_i$.
Note that the correlation functions displayed in Figs. \ref{eins} and
\ref{zwei} are not particularly significant for quantum information
processing; they were picked more or less arbitrarily to illustrate
our main result, the perfect transfer of many-qubit states.

To conclude, we have found an infinitely large class of
inhomogeneously coupled spin chain systems capable of perfect quantum
information transfer.  The freedom of choice within that class relaxes
the stringent constraints on the local spin couplings and magnetic
fields which previous models \cite{CDEL04,ACDE04} had to meet. We have
demonstrated perfect state transfer over fairly long distances in a
chain with almost homogeneous exchange coupling and without external
magnetic field. While many previous proposals have been restricted to
the transfer of single-spin states at zero temperature, the systems
discussed here are capable of transferring genuinely entangled states
involving several qubits, at arbitrary temperature.  Sensitivity to
perturbations like noise and imperfections \cite{BB05,RFB05,DRMF05}
will be the subject of further research.

\newcommand{\noopsort}[1]{}


\begin{thebibliography}{37}
\bibitem[{neg()}]{negative}
\bibinfo{note}{Negative
  $J_i$ may always be eliminated by appropriate $z$ spin axis rotations.}
\end{thebibliography}

\end{document}